\documentclass[11pt,a4paper]{article}

\usepackage{geometry}
\usepackage[utf8]{inputenc}
\usepackage{amssymb,amsfonts,amsmath,stmaryrd}
\usepackage[export]{adjustbox}
\usepackage{enumerate,float,indentfirst}
\usepackage{graphicx}
\usepackage{color}
\usepackage{tikz}
\usepackage{tikz-cd}
\usetikzlibrary{arrows,snakes,backgrounds}
\usepackage{url}
\definecolor{darkblue}{rgb}{0.1,0.1,.7}
\usepackage[colorlinks, linkcolor=darkblue, citecolor=darkblue, urlcolor=darkblue, linktocpage]{hyperref}
\usepackage{textcomp}
\usepackage{amsthm}
\usepackage{bbold}
\usepackage{mathtools}
\usepackage{multirow}
\usepackage{braket}
\usepackage{leftindex}
\usepackage[normalem]{ulem}
\usepackage[square, comma, compress,numbers]{natbib}
\usepackage{stackrel}
\usepackage{subcaption}
\usepackage{array}

\graphicspath{{figures/}}

\begin{document}

\vspace*{-.6in} \thispagestyle{empty}
\begin{flushright}
\end{flushright}
\vspace{1cm} {\Large
\begin{center}
{\bf Chaos in Systems with Quantum Group Symmetry}
\end{center}}
\vspace{1cm}
\begin{center}
{Victor Gorbenko, Aleksandr Zhabin}\\[2cm] 

\textit{Laboratory for Theoretical Fundamental Physics, Institute of Physics,\\ École Polytechnique Fédérale de Lausanne, Switzerland}\\
\vspace{1cm}

\vspace{1cm}
\end{center}

\vspace{4mm}

\begin{abstract}
Quantum groups have a long and fruitful history of applications in integrable systems. Can quantum group symmetries exist in the absence of integrability? We provide an explicit example of a system with quantum group global symmetry which is chaotic. The example is a spin chain with next-to-nearest interaction term. We show the chaotic behavior of the system by studying the Eigenvalue statistics. The spin chain is non-unitary but PT-symmetric and, in addition to chaos, exhibits an interesting transition after which the eigenvalues become complex. 

\end{abstract}
\vspace{.2in}
\vspace{.3in}
\hspace{0.7cm}

\section{Introduction}
Quantum group (QG) is a generalization of the usual groups which allows the commutation relation of the algebra elements to be non-linear. Initially quantum groups were discovered in connection with integrability, \cite{Takhtajan:1979iv,Kulish:1981dli}, see \cite{Faddeev:1996iy,Franchini:2016cxs} for reviews. To the best of our knowledge, all studies of physical models with the action of a QG focused on their integrable regimes, even though some of these systems also have range of parameters where they become chaotic \cite{Batchelor:1989uk,Read:2007ti,Ikhlef_2009}. Recently in \cite{paper1,paper2} it was emphasized that certain theories have global symmetries given by a QG, meaning that its charges commute with a Hamiltonian and states and operators fall into irreducible representations. In this case the number and type of constraints that QG imposes on the observables of  a theory is similar to that of a usual global symmetry and the number of integrals of motion is finite. This suggests that there should be models with QG symmetry that are not integrable. Nevertheless, the particular theory considered in \cite{paper1,paper2} as a main example is also integrable. In this paper we construct a model that is QG-invariant and manifestly chaotic. We do so by starting with QG-invariant version of the XXZ spin chain studied in \cite{Pasquier:1989kd} and deform it by an operator which preserves QG $U_q(sl_2)$ but breaks integrability.

Since the model is no longer integrable we solve it numerically using exact diagonalization. The resulting spectrum still contains degeneracies of eigenvalues in agreement with the QG symmetry, however, apart from these degeneracies, and modulo some discrete symmetries, the eigenvalues are distributed in a chaotic way. We demonstrate this by studying eigenvalue statistics which interpolates between the Poisson one, when the perturbation is turned off and the model is integrable, to a distribution typical of quantum chaos. The most distinctive feature of chaotic systems is the eigenvalue repulsion. An integrable Hamiltonian can likely have eigenvalues that are equal, or very close to each other, even if the states are not related to each other by any global symmetry. On the other hand, for a generic Hamiltonian the energy levels ``repel'' each other. The easiest way to see this is to consider a generic one-parameter family of two-by-two matrices. If matrices are Hermitian, this space is three-dimensional, while the subspace of matrices with the same eigenvalues has dimensionality one. This means that a generic curve will not intersect it. When the dimensionality of the Hilbert space becomes large, one can make precise statements about various statistical properties of eigenvalues of chaotic systems. These properties depend on various discrete symmetries, like parity and time-translation invariance. For Hermitian Hamiltonians the full classification was given in \cite{Altland:1997zz} and consists of 10 classes. 

It is now timely to mention that the theory we study in this paper is not unitary and the Hamiltonian is not Hermitian. Our choice is motivated by connection to the model studied in \cite{paper1,paper2} which is known to have a continuum limit when $q$ is on a unit circle. Instead of unitarity our model possesses the so-called PT-symmetry \cite{bender2024pt}. PT-symmetric models have recently attracted an increased interest due to a connection with open quantum systems, including their experimental realizations, see \cite{Ding:2022juv} and references therein. This symmetry also plays an important role in Lagrangian description of strongly-coupled CFTs as most recently discussed in \cite{Katsevich:2024jgq,Katsevich:2025ojk,ArguelloCruz:2025zuq}. Classification of non-unitary chaotic systems is much more complicated, and consists of as many as 38 classes \cite{bernard2002classification}, \cite{Kawabata:2018gjv}. While non-unitary systems are certainly more exotic, they can exhibit some interesting phenomena not present in the unitary ones. For example, as a function of the coupling constant our model undergoes a transition to the so-called PT-symmetry broken phase \cite{Bender_2007}, in which the eigenvalues are no longer real, but some of them come in complex-conjugate pairs. The co-existence of real and complex eigenvalues also makes the statistical properties much richer.

In the rest of the paper we review the integrable QG-invariant spin chain and introduce the perturbation under which the system exhibits strong evidence for the breakdown of integrability. Then we describe the procedure of modding-out the symmetries to identify a set of eigenvalues that  have well-defined statistical properties. In the appendix we focus on the low energy limit of our model and show that it flows to the same CFT as the undeformed integrable model. This implies that our perturbation is irrelevant in the RG sense. While we explicitly observe emergence of Virasoro symmetry in the IR, it does not appear to affect the chaotic properties of the rest of the spectrum.

\section{Spin chain with \texorpdfstring{$U_q(sl_2)$}{} symmetry and its algebraic structure}
\label{sec:PS_integrable}
In this section we define an integrable spin chain with $U_q(sl_2)$ symmetry and describe its algebraic structure following \cite{Pasquier:1989kd}. We will use this algebraic structure to construct non-integrable deformation in the following sections.

Let us start with defining the Hamiltonian of the antiferromagnetic spin chain of XXZ-type with particular boundary conditions, 
\begin{equation}\label{eq:PS_ham}
    \mathcal{H}_0 = \frac{1}{2} \sum_{i=1}^{L-1} \bigg( \sigma_{i}^{x}\sigma_{i+1}^{x} + \sigma_{i}^{y}\sigma_{i+1}^{y} + \frac{q+q^{-1}}{2} \sigma_{i}^{z}\sigma_{i+1}^{z} \bigg) - \frac{q-q^{-1}}{4} \big( \sigma^z_1 - \sigma^z_L \big)
\end{equation}
Such a choice of boundary conditions makes the Hamiltonian invariant under the $U_q(sl_2)$ symmetry.\footnote{Note that the sign in front of the boundary term is different from \cite{Pasquier:1989kd}. This does not affect neither the quantum group symmetry of the system nor the spectrum. However, we are using the ``$-$'' sign in order to match with the notations of \cite{paper1}.} We consider an even number of sites $L$. With the parametrization of $q$ being 
\begin{equation}\label{eq:q_param}
    q = e^{\frac{i\pi}{\mu+1}}, \qquad \mu \ge 1,
\end{equation}
the spin chain flows to a CFT in the IR, with the central charge $c=1-\frac{6}{\mu(\mu+1)}$. The spin chain \eqref{eq:PS_ham} is integrable and Bethe ansatz allows to find the spectrum. In the IR the spectrum was explicitly computed in \cite{Pasquier:1989kd}. For a more detailed discussion of the IR phase see Appendix \ref{app:IR_effects}.

The Hamiltonian is invariant under the $U_q(sl_2)$ action. The generators of $U_q(sl_2)$ on the lattice can be written as
\begin{equation}\label{eq:gen_sl2q_lattice}
    \begin{gathered}
        q^{H} = q^{\sigma^{z}} \otimes \dots \otimes q^{\sigma^{z}} \\
        E = \sum_{i=1}^{L} q^{-\sigma^{z}} \otimes \dots \otimes q^{-\sigma^{z}} \otimes \sigma_{i}^{+} q^{-\frac{\sigma^z}{2}} 
        \otimes 1 \otimes \dots \otimes 1 \\
        F = \sum_{i=1}^{L} 1 \otimes \dots \otimes 1 \otimes q^{\frac{\sigma^{z}}{2}} 
        \sigma_{i}^{-} \otimes q^{\sigma^{z}} \otimes \dots \otimes q^{\sigma^{z}}
    \end{gathered}
\end{equation}
Here $\sigma^{\pm}_i = \frac{1}{2}(\sigma_i^x \pm i\sigma_i^y)$. One can check that the generators commute with the Hamiltonian and provide the degeneracies of the spectrum. However, there is an important algebraic structure standing behind this Hamiltonian which makes the quantum group symmetry evident.

The algebraic structure that we need here is the Hecke algebra.\footnote{When the Hilbert space of the spin chain is made out of spin-1/2 representations of $U_q(sl_2)$ not all the elements of the Hecke algebra are linearly independent. For this reason the proper quotient is often considered, which is called the Temperley-Lieb (TL) algebra. The generators of TL algebra are denoted as $e_i$ and related to those of Hecke algebra as $e_i = q - R_i$. However, for our further discussion it is more convenient to work with the Hecke algebra.} It is generated by the elements $R_i$,
\begin{equation}\label{eq:R_i_explicit}
    R_{i} = \frac{1}{2} \bigg[ \sigma_{i}^{x} \sigma_{i+1}^{x} + \sigma_{i}^{y} \sigma_{i+1}^{y} + \frac{q+q^{-1}}{2} \big( \sigma_{i}^{z}\sigma_{i+1}^{z} + 1 \big) - \frac{q-q^{-1}}{2} \big( \sigma_{i}^{z} - \sigma_{i+1}^{z} - 2 \big) \bigg]\,,
\end{equation}
and all possible products of $R_i$, satisfying the following commutation relations:
\begin{equation}\label{eq:Hecke_rels}
    R_i R_{i+1} R_i = R_{i+1} R_i R_{i+1}, \qquad \qquad R_i^2 = (q-q^{-1}) R_i + 1.
\end{equation}
Note that the Hamiltonian can be rewritten is terms of the elements of the Hecke algebra as
\begin{equation}\label{eq:H0_Ris}
    \mathcal{H}_0 = -(L-1)\left( \frac{3q}{4} - \frac{1}{4q} \right) \mathbb{1} + \sum_{i=1}^{L-1} R_i\,.
\end{equation}
What is special about the Hecke algebra is that on the Hilbert space of the spin chain it is a commutant of $U_q(sl_2)$ \cite{Jones:1987dy}. That is, all the elements of the Hecke algebra commute with the quantum group.\footnote{One can think of this statement as a generalization of the standard Schur-Weyl duality between $GL(N)$ and symmetric group $S_n$ to the case of $q$-deformed algebras.} The fact that the Hamiltonian can be expressed in terms of the Hecke algebra generators then explains the quantum group symmetry, see \cite{paper1} for a review.

The Hamiltonian \eqref{eq:PS_ham} is not Hermitian for generic values of $q \in \mathbb{C}$: $\mathcal{H}_0 \ne \mathcal{H}_0^\dagger$. That is, generically the theory is not unitary. Nevertheless, for the values of $q$ on the unit circle, as it is in the parametrization \eqref{eq:q_param}, the Hamiltonian is known to be PT-symmetric \cite{Korff:2007qg}. By definition, it means that there exists a matrix $P$ such that $P \mathcal{H}_0 P^{-1} = \mathcal{H}_0^\dagger$. To show this, let us introduce a parity transformation $P$ which flips the spin chain around its center. It can be written as $P: \sigma_i^\alpha \to \sigma_{L-i+1}^\alpha$. Then the Hamiltonian \eqref{eq:PS_ham} satisfies,
\begin{equation}
\label{eq:PT_sym_0}
    P\mathcal{H}_0(q) P^{-1} = \mathcal{H}_0^T(q^{-1}) = \mathcal{H}_0^\dagger,
\end{equation}
where the last equality holds when $q$ is on the unit circle. This means that the Hamiltonian is adjoint under the $P$-metric and a conserved in time inner product can be defined \cite{Bender_2007,Bender:1998ke,Mannheim:2009zj}. The PT-symmetry implies also the constraint on the spectrum: the eigenvalues can be either real or  come in complex conjugated pairs. Sometimes it is called unbroken and broken PT-symmetry phases in the terminology of \cite{Bender_2007}. As it was rigorously shown in \cite{Morin-Duchesne:2015afa}, the spectrum of $\mathcal{H}_0$ is purely real for $q$ on the unit circle.

In addition to the PT-symmetry, the Hamiltonian is also symmetric under transposition,
\begin{equation}
\label{eq:transpose_0}
    \mathcal{H}_0 = \mathcal{H}_0^{T}.
\end{equation}
The formulas \eqref{eq:transpose_0} and \eqref{eq:PT_sym_0} specify the universality class of the complex Hamiltonian.\footnote{The symmetry class is BDI$^\dagger$ in the notations of \cite[Appendix B]{Kawabata:2018gjv}. } Since $\mathcal{H}_0$ is integrable, for any value of $q$ we expect the Poisson distribution for the level spacing.

\section{Next-to-nearest deformation preserving symmetries}
Now let us construct a deformation term that breaks the integrable structure of the Hamiltonian but preserves the $U_q(sl_2)$ symmetry. Generically, an integrable structure of spin chains can be broken by adding a next-to-nearest interaction term, see for example \cite{DAlessio:2015qtq,rabson2004crossover,gubin2012quantum,Rabinovici:2022beu}. For the spin chain \eqref{eq:PS_ham} we build up such a term quadratic in Hecke algebra generators \eqref{eq:R_i_explicit}. 

Let us start with writing the most general next-to-nearest interaction term quadratic in $R_i$:
\begin{equation}\label{eq:V_generic}
    V(\lambda) = \lambda \sum_{i=1}^{L-2} \bigg( a(q) R_{i}R_{i+1} + b(q) R_{i+1}R_{i} + c(q) R_{i} + d(q) R_{i+1} + e(q) \mathbb{1} \bigg)
\end{equation}
with unknown coefficients $a,b,c,d,e$ which can possibly depend on $q$. By construction, $V(\lambda)$ commutes with the $U_q(sl_2)$. 

Now, let us impose some restrictions on the coefficients $a,b,c,d,e$ by reproducing known results in the $q \to 1$ limit. In this limit we would like to match with the standard next-to-nearest deformation term of the XXX spin chain \cite{majumdar1969next,Eggert:1996er,White:1996mr},
\begin{equation}\label{eq:V_qto1}
    V(\lambda) \bigg|_{q \to 1} = \frac{\lambda}{2} \sum_{i=1}^{L-2} \bigg( \sigma_{i}^{x}\sigma_{i+2}^{x} + \sigma_{i}^{y}\sigma_{i+2}^{y} + \sigma_{i}^{z}\sigma_{i+2}^{z} \bigg).
\end{equation}
which preserves the $SU(2)$ symmetry and breaks integrability of the Heisenberg chain.\footnote{For small $\lambda$ such a deformation has been recently discussed only to weakly break integrability, see for example \cite{kurlov2022quasiconserved,orlov2023adiabatic,Surace:2023wqq}. That is, it is possible to construct an infinite set of quasi-conserved charges which commute up to $\mathcal{O}(\lambda^2)$. In this case the thermalization happens at later times and the level spacing statistics matches with the RMT predictions for a larger value of the coupling constant \cite{szasz2021weak}. However, we consider generic values of $\lambda$.} This matching restricts the unknown coefficients to be
\begin{equation}\label{eq:limit_abcde}
        a(1) = b(1) = 1, \qquad c(1) = d(1) = -1, \qquad e(1) = \frac{1}{2}.
\end{equation}

For generic $q$ we have a freedom in the choice of coefficients. Let us take
\begin{equation}
\begin{gathered}
    a(q) = b(q) = 1, \\
    c(q) = d(q) = - \bigg(q+\frac{q-q^{-1}}{2}\bigg), \\
    e(q) = (q^2-1) + \frac{q^2+q^{-2}}{4}.
\end{gathered}
\end{equation}
It is easy to check that in the limit $q \to 1$ they become exactly \eqref{eq:limit_abcde}. Finally, we write the next-to-nearest deformation term as,\footnote{In terms of Temperley-Lieb generators it can be equivalently written as
\begin{equation*}
    V(\lambda) = \lambda \sum_{i=1}^{L-2} \bigg( e_{i}e_{i+1} + e_{i+1}e_{i} - \frac{q+q^{-1}}{2} (e_{i} + e_{i+1}) + \frac{q^2+q^{-2}}{4} \mathbb{1} \bigg)\,.
\end{equation*}
This bulk deformation, although with different boundary conditions, was studied in \cite{Ikhlef_2009}. }\textsuperscript{,}\footnote{A certain different choice of coefficients might yield an integrable deformation. Higher spin conserved charge quadratic in Temperley-Lieb generators was considered in \cite{Miao:2022dau}.}
\begin{equation}
    V(\lambda) = \lambda \sum_{i=1}^{L-2} \bigg( R_{i}R_{i+1} + R_{i+1}R_{i} - \bigg[ q+\frac{q-q^{-1}}{2} \bigg] (R_{i} + R_{i+1}) + \bigg[ q^2-1 + \frac{q^2+q^{-2}}{4} \bigg] \mathbb{1} \bigg)\,.
\end{equation}
To efficiently perform numerical computations with this term, it is useful to have an expression for $V(\lambda)$ in terms of Pauli matrices,
\begin{equation}\label{eq:nnInt_pauli}
    V(\lambda) = \frac{\lambda}{2} \sum_{i=1}^{L-2} \bigg( \sigma_{i}^{x}\sigma_{i+2}^{x} + \sigma_{i}^{y}\sigma_{i+2}^{y} + \sigma_{i}^{z}\sigma_{i+2}^{z} + A_{i} + B_{i} \bigg)\,,
\end{equation}
where the terms $A_i$ and $B_i$ have simple expressions,
\begin{equation}
    \begin{gathered}
        A_{i} = \frac{(q-q^{-1})^{2}}{4} \bigg( \sigma_{i}^{z}\sigma_{i+1}^{z} + \sigma_{i+1}^{z}\sigma_{i+2}^{z} \bigg) - \frac{q^{2}-q^{-2}}{4} (\sigma_{i}^{z} - \sigma_{i+2}^{z})\,, \\
        B_{i} =  \frac{q-q^{-1}}{2} \bigg( \sigma_{i}^{x} \sigma_{i+1}^{x} \sigma_{i+2}^{z} + \sigma_{i}^{y} \sigma_{i+1}^{y} \sigma_{i+2}^{z} - \sigma_{i}^{z} \sigma_{i+1}^{x} \sigma_{i+2}^{x} - \sigma_{i}^{z} \sigma_{i+1}^{y} \sigma_{i+2}^{y} \bigg)\,.  \\
    \end{gathered}
\end{equation}
Here, $A_i$ is a diagonal matrix and $B_i$ has only off-diagonal non-zero matrix elements. Clearly, both $A_i$ and $B_i$ vanish as $q \to 1$.

From now on we are going to work with the Hamiltonian
\begin{equation}\label{eq:ham_final}
    \mathcal{H}(\lambda) = \mathcal{H}_0 + V(\lambda)
\end{equation}
By construction, it is invariant under $U_q(sl_2)$ symmetry. Let us also note that the symmetries \eqref{eq:PT_sym_0} and \eqref{eq:transpose_0} are also preserved for any $\lambda$,
\begin{equation}\label{eq:pt_total}
\begin{gathered}
    \mathcal{H}^T = \mathcal{H}\,, \\
    P \mathcal{H}(q) P^{-1} = \mathcal{H}^T(q^{-1}) = \mathcal{H}^\dagger,
\end{gathered}
\end{equation}
which implies that the Hamiltonian stays in the same universality class after the deformation. In particular, it means that after the deformation we have the same constraint on the eigenvalues: they can be either real or come in complex conjugated pairs. In contrast to the undeformed Hamiltonian $\mathcal{H}_0$, the eigenvalues of $\mathcal{H}(\lambda)$ are complex for generic values of parameters $q$ and $\lambda$. As we will discuss later, for some values of the parameters it is nevertheless possible to get a purely real spectrum.

\section{Identifying global symmetries and choosing subsector}
In order to analyze the eigenvalue statistics of the Hamiltonian \eqref{eq:ham_final}, we first have to identify all the global symmetries of the system and then project onto the subsector with the same quantum numbers under all global internal symmetries. The states with different quantum numbers are generally uncorrelated and do not represent the universal integrable or chaotic behavior.

\begin{enumerate}
    \item \textbf{Quantum group symmetry.} As already stated above, the Hamiltonian $\mathcal{H}(\lambda)$ is invariant under $U_q(sl_2)$. All the eigenstates of the Hamiltonian have quantum numbers $\{\ell,m\}$ under this symmetry, where $\ell \in \mathbb{N} \cup \{0\}$ (there are no half-integer spins) and $m \in \{-\ell,\ldots, \ell\}$.\footnote{{To be precise, such a Hilbert space decomposition is true only when $q$ is not a root of unity. In the case when $q$ hits a root of unity, reducible but indecomposable representation of $U_q(sl_2)$ might appear. In the following numerical analysis the values of $(\log q)/i\pi$ cannot be taken irrational because of finite numerical precision. However, they are taken to be generic enough, such that for studied length $L$ of the spin chain these representations do not appear.}} The quantum number $m$ is measured by an operator
    \begin{equation}
        S^z = \frac{1}{2} \sum_{i=1}^{L} \sigma^z_i.
    \end{equation}
    For a given state in the $\sigma^z$ basis, $S^z$ is equal to the total number of spins up minus total number of spins down. 
    
    Let us consider a subsector of the Hilbert space with fixed $S^z$. It contains only a single state $\{\ell,S^z\}$ from each quantum group multiplet with $\ell \ge S^z$. The subsector is bigger than needed since we want to restrict to a subsector where all the states have the same $\ell$ and $m$. Let us denote the set of eigenvalues of $\mathcal{H}$ in this subsector as $\mathcal{E}_{S^z}$. To further restrict the set of eigenvalues one can additionally compute the eigenvalues of $\mathcal{H}$ in the $S^z+1$ subsector, denoted by $\mathcal{E}_{S^z+1}$. All the states in this subsector have $\ell \ge S^z+1$. By the $U_q(sl_2)$ symmetry all the eigenvalues in $\mathcal{E}_{S^z+1}$ are already contained in $\mathcal{E}_{S^z}$. Then, by subtracting $\mathcal{E}_{S^z+1}$ from $\mathcal{E}_{S^z}$ one can get exactly the set of eigenvalues labeled by $\{\ell=S^z,m=S^z\}$. In what follows we restrict ourselves to a subsector $S^z=1$ to avoid possible complications that may arise in the $S^z=0$ subsector, such as a presence of spin-flip operation accompanied by $q \to q^{-1}$ \cite{Quella:2020nmv,Franke:2024vdo}.

    \item \textbf{q-Parity symmetry.} Besides the quantum group symmetry, there is an additional global spacetime symmetry in the system. The Hamiltonian $\mathcal{H}(\lambda)$ commutes with the following operator,
    \begin{equation}\label{eq:def_q_parity}
        P_q := P^{(q)}_{1,L} P^{(q)}_{2,L-1} \ldots P^{(q)}_{\frac{L}{2},\frac{L}{2}+1}\,,
    \end{equation}
    where the operators $P^{(q)}_{i,j}$ are introduced in terms of Hecke algebra generators \eqref{eq:R_i_explicit} in the following way. First, if $j=i+1$ then we define,
    \begin{equation}\label{eq:def_nn_transpos}
        P^{(q)}_{i,i+1} := R_i\,.
    \end{equation}
    Note that when $q \to 1$ then $R_i$ becomes just a standard transposition matrix:
    \begin{equation}\label{eq:limit_Ri}
        R_i \bigg|_{q \to 1} = P^{(1)}_{i,i+1} = \mathbb{1}_{i-1} \otimes \begin{pmatrix}
            1 & 0 & 0 & 0\\
            0 & 0 & 1 & 0\\
            0 & 1 & 0 & 0\\
            0 & 0 & 0 & 1
        \end{pmatrix} \otimes \mathbb{1}_{L-i-1}
    \end{equation}
    With the formula \eqref{eq:def_nn_transpos} we now define the operators $P^{(q)}_{i,j}$ as,
    \begin{equation}
        P^{(q)}_{i,j} := P^{(q)}_{i,i+1} P^{(q)}_{i+1,i+2} \ldots P^{(q)}_{j-1,j} P^{(q)}_{j-2,j-1} \ldots P^{(q)}_{i,i+1}\,.
    \end{equation}
    For generic $q$, the formula for $P^{(q)}_{i,j}$ completes the definition of the operator \eqref{eq:def_q_parity}. In the limit $q\to1$ the operator $P^{(1)}_{i,j}$ becomes a product of nearest-neighbour transpositions, which results into a transposition of lattice sites $i$ and $j$. The operator $P_q$ in the limit $q\to1$ becomes accordingly just the standard parity operator $P$ which flips the spin chain around its center. That is why we will refer to $P_q$ as a $q$-deformed parity operator.\footnote{This symmetry can be viewed as a natural $q$-deformation of the parity symmetry. The usual parity symmetry is present in the standard open XXX or XXZ spin chains. To study the statistics of eigenvalues in such systems one has to choose the subsector with all eigenvectors of the Hamiltonian being either even or odd under the parity.} In particular, one can check that its action on $R_i$-s also looks like a flip of the spin chain around its center,
    \begin{equation}
        P_q R_i P_q^{-1} = R_{L-i}.
    \end{equation}
    This action manifestly makes the $q$-parity operator commute with the undeformed Hamiltonian \eqref{eq:H0_Ris}. For the deformation term it is the choice of coefficients $a(q)=b(q)$ and $c(q)=d(q)$ in \eqref{eq:V_generic} which preserves the symmetry. Since it is constructed as a product of Hecke algebra elements, it also commutes with the QG:
    \begin{equation}
        [P_q, \mathcal{H}] = 0, \qquad [P_q, U_q(sl_2)] = 0.
    \end{equation}
    
    Note that the $q$-parity operator is highly non-local on the spin chain since it contains the product of all local lattice operators $R_i$. Using only the first of the commutation relations \eqref{eq:Hecke_rels} of the Hecke algebra and the formula \eqref{eq:def_nn_transpos}, one can derive the closed form expression for $P_q$,
    \begin{equation}\label{eq:Pq_explicit}
        P_q = \bigg( R_1 \ldots R_{L-1} \bigg) \bigg( R_1 \ldots R_{L-2} \bigg) \ldots \bigg( R_1 R_2 \bigg) R_1\,.
    \end{equation}

    The $q$-parity operator further splits the spectrum into subsectors with different quantum numbers under its action. The numerical diagonalization of $P_q$ strongly supports that the subsector of $\mathcal{H}$ with quantum numbers $\{\ell,m\}$ under $U_q(sl_2)$ symmetry has the following eigenvalues under the action of $P_q$, 
    \begin{equation}\label{eq:eigenv_Pq}
        \pm q^{\ell(\ell+1) + \frac{L^2}{4} - L}.
    \end{equation}
    When $q \to 1$, the ``$\pm$'' sign corresponds to usual parity even and odd states respectively. By analogy, let us call the eigenstates of $P_q$ even and odd under the $q$-parity. In what follows we would like to study only the even $q$-parity eigenstates. Unlike the standard parity operator $P$, analytical projection to the $q$-parity even states is much harder. Because of that issue we need to come up with another algorithm of selecting $q$-parity even eigenstates. We discuss the algorithm in the next section.

    A different choice of coefficients in the interaction, namely $a \ne b, c \ne d$ can break the $q$-parity symmetry, which would simplify the computation of the eigenvalues in the chaotic regime. On the other hand, in this case the number of eigenvalues belonging to the same symmetry sector will change discontinuously as compared to the integrable case and hence it would obscure our diagnostics of the transition to chaotic behavior.

\end{enumerate}

\section{Eigenvalue statistics}
\subsection{Procedure of projecting out global symmetries}
\label{sec:procedure}
Let us start with describing the procedure of extracting the $q$-parity even states in the subsector with quantum numbers $\{\ell=1,m=1\}$ under the $U_q(sl_2)$ symmetry. As the expression for $P_q$ is highly non-local \eqref{eq:Pq_explicit} and an analytic form of its eigenstates is unknown, many numerical operations with it take a lot of time. The procedure is chosen in order to optimize the computational time.
\begin{itemize}
    \item First, we construct the operators $\mathcal{H}$ and $P_q$ directly in a subsector with $S^z = 1$. Let us denote these matrices as $\mathcal{H}^{(1)}$ and $P_q^{(1)}$. This is possible to do given the explicit expressions \eqref{eq:PS_ham}, \eqref{eq:nnInt_pauli} in terms of Pauli matrices for $\mathcal{H}$ and the expressions \eqref{eq:R_i_explicit}, \eqref{eq:Pq_explicit} for $P_q$. Since both of these operators are block diagonal in the $S^z$ basis, we can compute their product directly in this subsector: $P_q^{(1)} \mathcal{H}^{(1)} = (P_q \mathcal{H})^{(1)}$.
    \item Second, by diagonalizing the matrix
    \begin{equation}\label{eq:pqtimesH}
        \left(P_q^{(1)} + q^{2 + \frac{L^2}{4} - L} \cdot \mathbb{1} \right) \mathcal{H}^{(1)}
    \end{equation}
    we obtain the set of eigenvalues
    \begin{equation}
        \mathcal{E} = \bigg\{ \left( \lambda_i + q^{2 + \frac{L^2}{4} - L} \right) E_i \bigg\}
    \end{equation}
    where $E_i$ are the eigenvalues of $\mathcal{H}^{(1)}$ and $\lambda_i$ are the eigenvalues of $P_q^{(1)}$ given by \eqref{eq:eigenv_Pq}. Note that this set of numbers corresponds to all the states with $m=1$ and $\ell \ge 1$. For all the $q$-parity odd eigenstates with $\ell=1$ we have $\lambda_i=-q^{2 + \frac{L^2}{4} - L}$, that is, all such states correspond to zeroes in the set $\mathcal{E}$. We remove all the zeroes and divide the rest by $2q^{2 + \frac{L^2}{4} - L}$. Let us call the new set $\mathcal{E}'$,
    \begin{multline}\label{eq:eprime_into_two}
        \mathcal{E}' = \left\{ \frac{\left( \lambda_i + q^{2 + \frac{L^2}{4} - L} \right)}{2q^{2 + \frac{L^2}{4} - L}} E_i \right\} =\\
        =\bigg\{ E_i \bigg\}_{\{\ell=1,m=1\}}^{\{q\text{-even}\}} \;\;\; \bigcup \;\;\; \left\{ \frac{\left( \lambda_i + q^{2 + \frac{L^2}{4} - L} \right)}{2q^{2 + \frac{L^2}{4} - L}} E_i \right\}_{\{\ell\ge2,m=1\}}
    \end{multline}
    For the $q$-parity even states with $\ell=1$ the prefactor in front of the energies disappears and we get the genuine set of energies $E_i$. For all the other states with $\ell \ge 2$ the prefactor remains and is a generic complex number for generic $q$. This allows to write $\mathcal{E}$ as two subsets as in \eqref{eq:eprime_into_two}.
    \item Finally, we diagonalize $\mathcal{H}^{(1)}$ and obtain the set of its eigenvalues $\mathcal{G} = \big\{ E_i \big\}$. These eigenvalues have quantum numbers $m=1$, $\ell \ge 1$, both $q$-parity even and odd. The only identical numbers in the two sets $\mathcal{E}'$ and $\mathcal{G}$ are those $q$-parity even eigenstates with $\ell=1,m=1$. This is exactly the desired set of eigenvalues.
\end{itemize}

\subsection{Analysis of different types of eigenvalues}
\label{sec:analysis}
Now we have all the ingredients to study the eigenvalue statistics of the Hamiltonian \eqref{eq:ham_final} with the next-to-nearest interaction term and $U_q(sl_2)$ symmetry. We consider eigenvalues from a subsector of $q$-parity even states with quantum numbers $\{\ell=1,m=1\}$.

The symmetry constraints \eqref{eq:pt_total} imply that the Hamiltonian $\mathcal{H}(\lambda)$ in principle has to be compared with a special ensemble of PT-symmetric matrices with complex entries, that are also symmetric under transposition. However, we compare the eigenvalue statistics in different regimes against the known Gaussian ensembles.

In the case when all the eigenvalues are real we study the level spacing distribution and the distribution of ratios of level spacings on the real line. That is, for an ordered set of energies $\big\{ E_i \big\}$ we compute the differences of nearest energies, $s_i = E_i - E_{i+1}$, and the ratio, $s_i/s_{i+1}$. To study the integrable vs chaotic behavior of the level spacing distribution, $P(s)$, we compare it with the Poisson vs Wigner distribution for the Gaussian Orthogonal Ensemble (GOE),
\begin{equation}\label{eq:PW_GOE}
    P_{\text{P}}(s) = e^{-s}, \qquad \qquad P_{\text{W}}(s) = \frac{\pi s}{2} e^{-\frac{\pi s^2}{4}}\,.
\end{equation}
Even though the Hamiltonian \eqref{eq:ham_final} is not from the GOE ensemble, it has the symmetry $\mathcal{H}^T = \mathcal{H}$. As we will see later, for the case of real eigenvalues it is nevertheless a good approximation of the distribution in the most chaotic regime. Let us also mention technical details of computing the distribution. While computing the real level spacing distribution we first drop the low-energy and high-energy tails of the energy spectrum, which correspond to the finite-size effects. We also perform the procedure of local unfolding \cite{mehta2004random,Atas:2013gvn} to go from a model-dependent distribution to the one predicted by random matrix universality.
    
As for the ratio distribution, it is convenient to define $\Tilde{r}_i = \min \left( \frac{s_i}{s_{i+1}}, \frac{s_{i+1}}{s_i} \right)$, and to study the distribution $P(\Tilde{r})$ \cite{Oganesyan:2007wpd,Atas:2013gvn}, see also \cite{Rabinovici:2022beu}. This distribution has an advantage compared to the level spacing distribution because it does not require local unfolding, which may be an ambiguous procedure. The distribution $P(\Tilde{r})$ for Poissonian and Wigner statistics is very well approximated by
\begin{equation}\label{eq:PW_GOE_ratio}
    P_{\text{P}}(\Tilde{r}) = \frac{2}{(1+\Tilde{r})^2}\,, \qquad \qquad P_{\text{W}}(\Tilde{r}) = \frac{27}{4}\frac{ \left( \Tilde{r}+\Tilde{r}^2 \right)}{ \left( 1+\Tilde{r}+\Tilde{r}^2 \right)^{5/2} }\,,
\end{equation}
for the GOE ensemble.

For the case of complex eigenvalues the above analysis is no longer applicable.\footnote{In principle, one can still study the statistics of purely real eigenvalues. Given that the constraints \eqref{eq:pt_total} imply localization of at least some part of the spectrum on the real line, one can still make statements about universal real level spacing distribution. See for example \cite{Xiao:2022gje} for this discussion, which also includes the universality class \eqref{eq:pt_total}. However, in practice the number of real eigenvalues in the given sector is not large enough to collect valid statistics.} In such cases it is more convenient to analyze the distribution of eigenvalues in the complex plane. We use a complex level spacing ratio (CLSR) defined as \cite{Sa:2020fpf},
\begin{equation}\label{eq:complex_spacing}
    z_i = \frac{E_i^{\rm NN}-E_i}{E_i^{\rm NNN}-E_i} \equiv r_i e^{i\theta_i},
\end{equation}
where for a given eigenvalue $E_i$ one has to find its nearest neighboring eigenvalue $E_i^{\rm NN}$ in the complex plane as well as its next-to-nearest neighboring eigenvalue. By definition $z$ is bounded in the unit circle, $|z| \le 1$. Now one can study a distribution $P(z)$ in the complex plane. The repulsion of complex eigenvalues in the chaotic regime manifests itself in the CLSR distribution as well. 
See for example \cite{Sa:2020fpf} for the plots of complex spacing ratio for generic complex matrices, so called Ginibre Unitary Ensemble (GinUE). An important quantity that we compute out of the distribution of $z_i$ is their average absolute value $\braket{r}$. This single number is used to distinguish between integrable and chaotic behavior. An advantage of $\braket{r}$ is that it can be computed both for the complex spectrum and for the real spectrum.\footnote{Note that when computed for the real spectrum, the average value of $\braket{r}$ does not have to coincide with average real spacing ratio $\braket{\Tilde{r}}$.} That is, this analysis is applicable for all values of $q$ and $\lambda$. The value of $\braket{r}$ for different ensembles is presented in the Table \ref{tbl:average_r}. Matrices from GinUE ensemble do not satisfy the constraints \eqref{eq:pt_total}, but we use this ensemble as a reference mark when studying complex eigenvalues. We will see that by varying the parameters $q$ and $\lambda$, the spectrum interpolates between GOE and GinUE predictions. Detailed discussion is presented in section \ref{sec:complex_analysis} around figure \ref{fig:average}.\footnote{Since time-reversal symmetry (TRS) does not get broken in our model when complex eigenvalues appear, one might want to compare the average value $\braket{r}$ with that of Ginibre Orthogonal Ensemble (GinOE). However, both GinUE and GinOE share the same universal spectral statistics \cite{Hamazaki:2020kbp}, since TRS introduces correlations only between complex-conjugate eigenvalue pairs and leaves nearest-neighbor statistics unaffected. Furthermore, GinOE matrices are real and not symmetric, which is again a different ensemble than the one given by \eqref{eq:pt_total}.}
\begin{table}[h!]
\centering
\renewcommand{\arraystretch}{2}
\begin{tabular}{|>{\centering\arraybackslash}p{2cm}|>{\centering\arraybackslash}p{2cm}|>{\centering\arraybackslash}p{2cm}|>{\centering\arraybackslash}p{2cm}|}
\hline
   Ensemble & Poisson & GOE   & GinUE \\
\hline
 $\braket{r}$ & 0.500   & 0.570 & 0.738 \\
\hline
\end{tabular}
\caption{Average absolute value of CLSR computed for different ensembles.}
\label{tbl:average_r}
\end{table}

\subsection{Numerical results}
Let us now present the eigenvalue statistics for different parameters $q$ and $\lambda$. We work with the length $L=16$ of the spin chain, the total dimension of the Hilbert space is $2^{16}$. According to the procedure described in section \ref{sec:procedure} we build and diagonalize the matrices in the subsector $S^z=1$ of the size $\binom{16}{8+1} = 11440$. The first set of $11440$ eigenvalues are those of $\mathcal{H}^{(1)}$. Then we calculate the eigenvalues of the matrix \eqref{eq:pqtimesH} and remove zeroes, working with $10^{-6}$ precision. After removing zeroes, 9696 eigenvalues remain in the set $\mathcal{E}'$. We compare the set of eigenvalues $\mathcal{E}'$ with those of $\mathcal{H}^{(1)}$ with $10^{-6}$ precision. After completing the procedure, the number of eigenvalues in the $q$-parity even subsector with $\{\ell=1,m=1\}$ is equal to $1688$. 

The final number of $1688$ eigenvalues can be verified through a combinatorial argument. In the computational basis, every basis vector is encoded by a binary sequence of length $L=16$, with spin-up $(1)$ or spin-down $(0)$ at each lattice site. One needs to find the number of sequences in a given $S^z$ sector, that are symmetric under reflection around the center. These would correspond to (usual) parity even basis eigenvectors. There are exactly $5720$ of such vectors in $S^z=1$ sector, and $4032$ of those in $S^z=2$ sector. Restricting to a single irreducible sector under non-abelian symmetry, namely $\ell=1,m=1$, requires subtracting the latter from the former, resulting in exactly $1688$ eigenvalues. Note that when working with $SU(2)$ invariant spin chain, one could diagonalize the Hamiltonian directly in this parity even basis. However, the eigenvectors of q-parity operator $P_q$ are not known to be constructed analytically, and the procedure of section \eqref{sec:procedure} is needed. Nevertheless, from the formula \eqref{eq:eigenv_Pq} we know that the number of eigenvalues stays the same. This independent check supports the choice of numerical precision: (a) no extra eigenvalues from sectors with $\ell\ge2$ contaminate the final set of eigenvalues; (b) no eigenvalues from the sector we are interested in were accidentally removed.

For all the considered values of $q$, when taking $\lambda = 0$ the Hamiltonian $\mathcal{H}(0) \equiv \mathcal{H}_0$ is integrable, all the eigenvalues are real, and the expectation is that real Poisson distribution describes the level spacing. 

\subsubsection{Real spectrum}
\label{sec:real_spectrum}
Let us first consider the case when the parameter $q$ is close to $1$. Let us take it to be
\begin{equation}
    q = e^{\frac{i\pi}{10.4}}
\end{equation}
From the explicit expressions \eqref{eq:PS_ham}, \eqref{eq:nnInt_pauli} one can expect that for this value of $q$ the spectrum of the Hamiltonian \eqref{eq:ham_final} would not change much compared to the spectrum of XXX chain with the next-to-nearest interaction term \eqref{eq:V_qto1}. Indeed, what we find is that the spectrum in our sector is purely real. This allows us to perform the standard real analysis described in section \ref{sec:analysis}. 

\begin{figure}[t]
\subfloat{{\includegraphics[width=5cm,valign=c]{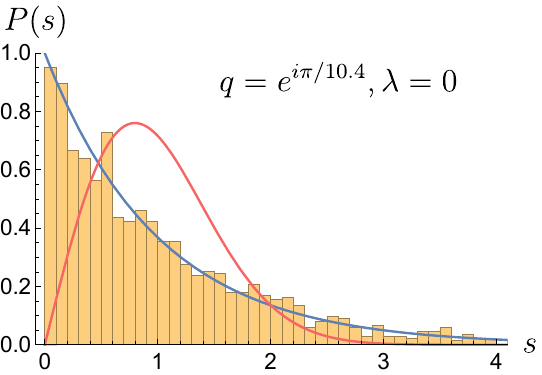} }}
\quad
\subfloat{{\includegraphics[width=5cm,valign=c]{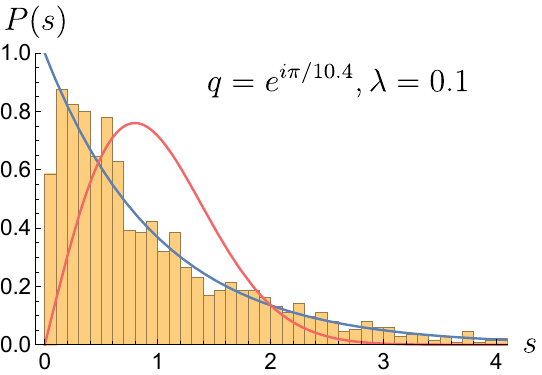} }}
\quad
\subfloat{{\includegraphics[width=5cm,valign=c]{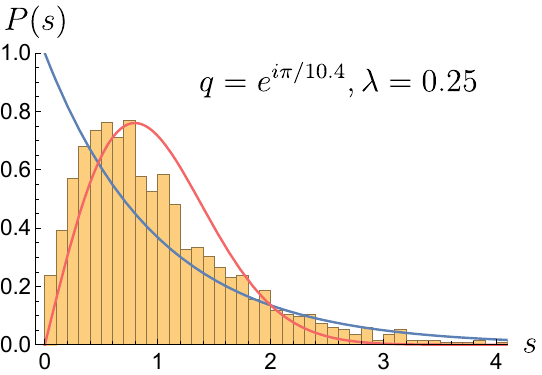} }}
\quad
\subfloat{{\includegraphics[width=5cm,valign=c]{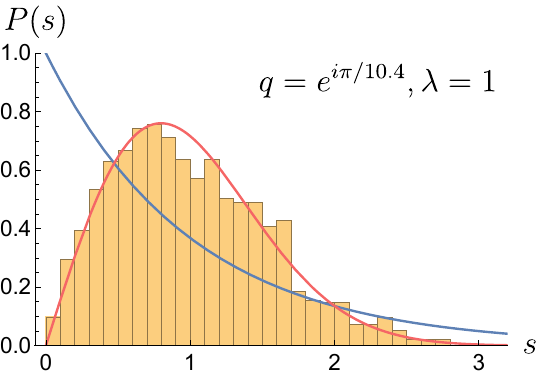} }}
\quad
\subfloat{{\includegraphics[width=5cm,valign=c]{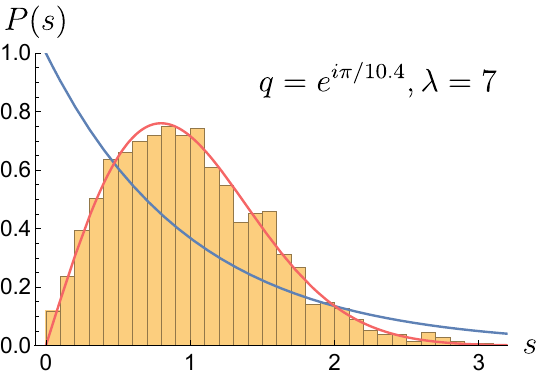} }}
\quad
\subfloat{{\includegraphics[width=5cm,valign=c]{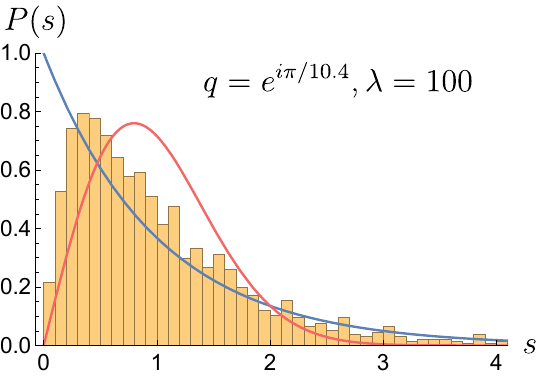} }}
\caption{Level spacing distribution $P(s)$ for real eigenvalues from the $q$-parity even sector with $\{\ell=1,S^z=1\}$ with fixed parameters $L=16,q=e^{i\pi/10.4}$ and different $\lambda$. The blue and red solid lines are respectively Poissonian and Wigner distributions \eqref{eq:PW_GOE} for the GOE ensemble.}
\label{fig:l16q10d4withLocUnfold}
\end{figure}

The plots for the real level spacing distribution for different values of $\lambda$ are presented in the Figure \ref{fig:l16q10d4withLocUnfold}.\footnote{As it was mentioned in section \eqref{sec:analysis}, to obtain such distributions from original set of eigenvalues we perform two additional steps. (1) We drop the $10\%$ of the spectrum on each side (the smallest and the highest eigenvalues). (2) We perform the local unfolding procedure: the list of eigenvalue spacings is divided into consecutive blocks of 20 elements, each block is normalized by its mean value. The length of the block is chosen large enough to reliably estimate the local mean value and small enough such that the density does not vary significantly within the block. Such a procedure allows to compare physical spectrum with universal RMT predictions.} For each $\lambda$ we compare the distribution with the Poissonian and Wigner distributions \eqref{eq:PW_GOE} represented by the solid blue and red lines respectively. First, as expected, we see that for $\lambda=0$ the level spacing distribution is well approximated by the Poisson distribution, which agrees with $\mathcal{H}_0$ being integrable. Second, as we increase the coupling constant $\lambda$ the distribution shifts, which indicates the appearance of level repulsion in the system. When $\lambda$ reaches values of order 1 the level spacing distribution is well approximated by the Wigner surmise for the GOE ensemble. For these values of $\lambda$ the level repulsion is the strongest and the system is in the most chaotic regime.\footnote{Strictly speaking, the system becomes chaotic for any value of $\lambda>0$. However, to see the maximal level repulsion the coupling constant has to be of the order of the average energy spacing in the Hamiltonian. That is, one can observe the maximally chaotic distribution for smaller values of $\lambda$ by increasing the system size $L$ \cite{DAlessio:2015qtq}. } Even though the Hamiltonian $\mathcal{H}(\lambda)$ with the symmetries \eqref{eq:pt_total} is not from the GOE ensemble, we see a good matching with the GOE statistics. This is due to the fact that the Hamiltonian is still symmetric under transposition, $\mathcal{H}^T = \mathcal{H}$, and all the eigenvalues are real. That is, the Hamiltonian is in the so-called PT-symmetry unbroken phase, and the second equation in \eqref{eq:pt_total} does not affect the eigenvalue statistics. Third, as we keep increasing $\lambda$ the distribution tends to shift back to the Poissonian curve. It is an expected behavior for $q=1$, because for large $\lambda$ one can neglect the $\mathcal{H}_0$ term in the Hamiltonian and treat the next-to-nearest deformation term \eqref{eq:V_qto1} as a system of two decoupled integrable spin chains on even and odd sites respectively. For $q\ne1$ we do not know if the term \eqref{eq:nnInt_pauli} is integrable by itself, nevertheless, since $q$ is close to $1$ we observe that the level spacing for large $\lambda$ is close to the integrable distribution.

\begin{figure}[h!]
\subfloat{{\includegraphics[width=5cm,valign=c]{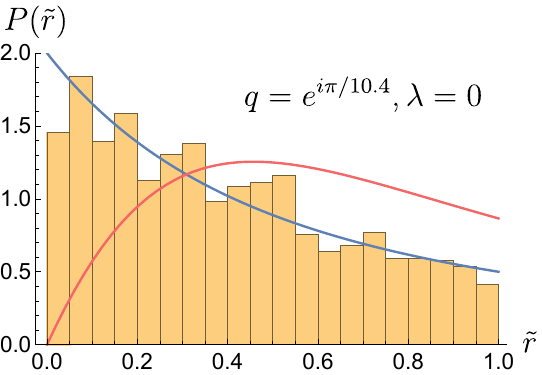} }}
\quad
\subfloat{{\includegraphics[width=5cm,valign=c]{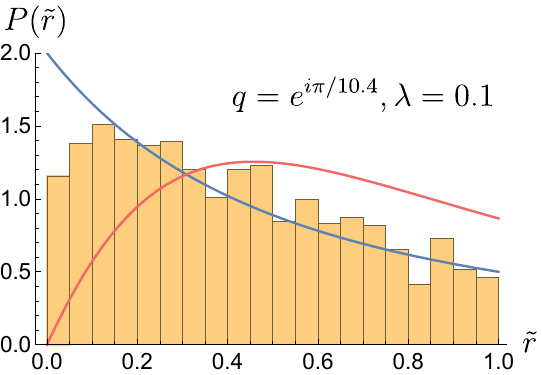} }}
\quad
\subfloat{{\includegraphics[width=5cm,valign=c]{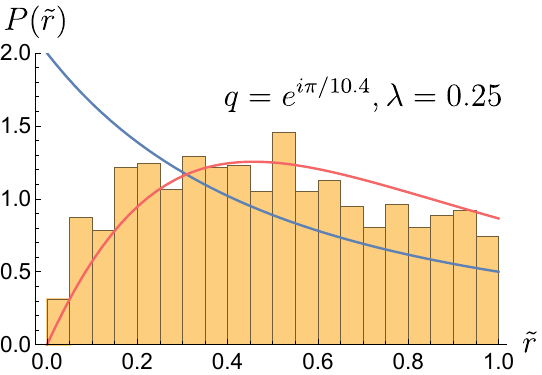} }}
\quad
\subfloat{{\includegraphics[width=5cm,valign=c]{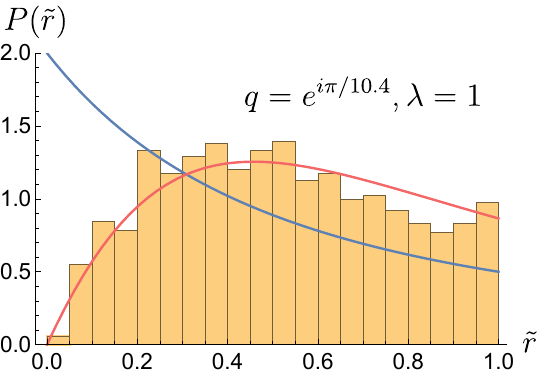} }}
\quad
\subfloat{{\includegraphics[width=5cm,valign=c]{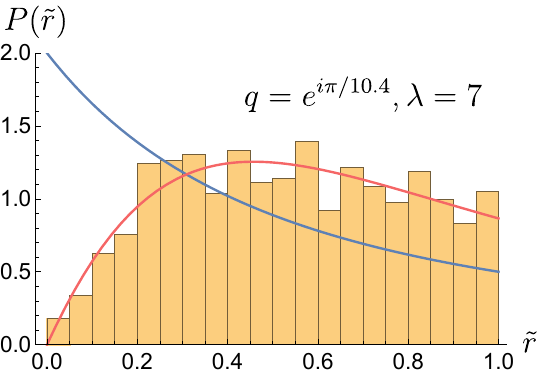} }}
\quad
\subfloat{{\includegraphics[width=5cm,valign=c]{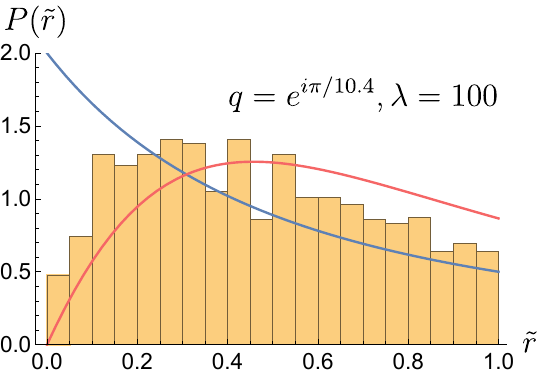} }}
\caption{Spacing ratio distribution $P(\Tilde{r})$ for real eigenvalues from the $q$-parity even sector with $\{\ell=1,S^z=1\}$ with fixed parameters $L=16,q=e^{i\pi/10.4}$ and different $\lambda$. The blue and red solid lines are respectively Poissonian and Wigner distributions approximating ratio distributions \eqref{eq:PW_GOE_ratio} for the GOE ensemble.}
\label{fig:l16q10d4ratiodistr}
\end{figure}

As explained above, for real eigenvalues we also plot the ratio distribution of level spacings, $P(\Tilde{r})$, see Figure \ref{fig:l16q10d4ratiodistr}. For each value of $\lambda$ we compare the distributions with the Poissonian and Wigner distributions \eqref{eq:PW_GOE_ratio}. The behavior of the ratio distribution with varying $\lambda$ is similar to the behavor of level spacing distribution. For $\lambda=0$ it is well approximated by Poisson curve. As $\lambda$ increases the distribution shifts to the Wigner curve for the GOE ensemble. And for large $\lambda$ the distribution tends back to the integrable one.

Both distributions of level spacings $P(s)$ given in Figure \ref{fig:l16q10d4withLocUnfold} and of ratios of level spacings $P(\Tilde{r})$ given in Figure \ref{fig:l16q10d4ratiodistr} serve as a strong indicator that the next-to-nearest interaction term breaks integrability of the system.

\subsubsection{Complex spectrum}\label{sec:complex_analysis}
We now move on to analysis of the spectrum for different values of $q$. In particular, we want to explore values of $q$ on the unit circle given by \eqref{eq:q_param} for which we have an analytic control in the continuum limit of $\mathcal{H}_0$. The first observation which one can make when moving $q$ further away from $1$ (equivalently, moving $\mu$ further away from $\infty$) is that complex eigenvalues start to appear. By the symmetry \eqref{eq:pt_total} they appear in complex conjugated pairs. For generic $q$ and $\lambda$ the majority of eigenvalues become complex, that is, only real part of the spectrum does not contain enough statistics. To make the full use of the eigenvalues in a given subsector, we examine them with the complex analysis of section \eqref{sec:analysis}.

\begin{figure}[t]
    \centering
    \includegraphics[width=0.8\linewidth]{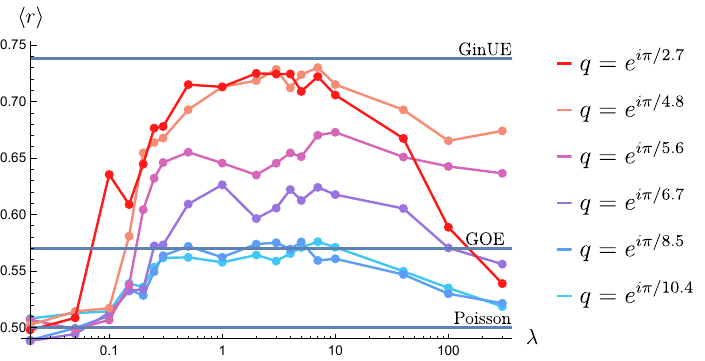}
    \caption{Value of average CLSR, $\braket{r}$, as a function of $\lambda$ for different values of parameter $q$. Solid horizontal lines are the values of $\braket{r}$ for different ensembles taken from Table \ref{tbl:average_r}. For the values of $q$ close to 1 all the eigenvalues in the maximally chaotic regime are real and the values of $\braket{r}$ approach GOE ensemble. For $q$ far away from 1 almost all the eigenvalues become complex in chaotic regime and $\braket{r}$ reaches the values comparable with GinUE ensemble.}
    \label{fig:average}
\end{figure}

For different values of $q$ and $\lambda$ we compute the complex level spacing ratio \eqref{eq:complex_spacing} and then compute an average absolute value $\braket{r}$. The plot of absolute CLSR as a function of $\lambda$ for several different values of $q$ is presented in the Figure \ref{fig:average}. There are several important observations that one can make from this plot. Foremost, for all values of $q$ CLSR starts around $0.5$ for $\lambda=0$, which corresponds to an integrable regime. Then, the value of $\braket{r}$ increases as a function of $\lambda$ and reaches some plateau when the coupling constant becomes of order $1$, that is, the system reaches maximally chaotic phase. The value of the plateau depends on how many complex eigenvalues appear in the spectrum for a given $q$. For example, for $q=e^{i\pi/10.4}$, when all the eigenvalues are real, the behavior of $\braket{r}$ is consistent with the analysis of section \ref{sec:real_spectrum}: in the maximally chaotic regime average CLSR reaches the value for the GOE ensemble. For different values of $q$, the further away $q$ is from $1$ the more complex eigenvalues appear in the spectrum. See Table \ref{tbl:max_complex} for the maximal number of complex eigenvalues, $\mathcal{N}_{\mathbb{C}}$, that appears in the spectrum for a fixed $q$ as the coupling constant reaches the maximally chaotic regime. One can observe a clear correlation between $\mathcal{N}_{\mathbb{C}}$ and the value of the plateau for $\braket{r}$ in the Figure \ref{fig:average}. By tuning the parameter $q$ one can then transition between unbroken and broken PT-symmetric phases of the Hamiltonian in the terminology of \cite{Bender_2007}, and this transition is captured by the values of $\braket{r}$. A similar phenomenon was observed in \cite{Sharma:2024fqc} where complex eigenvalue statistics of a different PT-symmetric system was studied. Lastly, we observe that maximal values of $\braket{r}$ for $q=e^{i\pi/2.7}$ and $q=e^{i\pi/4.8}$ become comparable with GinUE ensemble, but do not exactly reach its value. Recall that the Hamiltonian \eqref{eq:ham_final} is from the PT-symmetry class \eqref{eq:pt_total}, so average CLSR does not have to be the same as for GinUE matrices, which do not have any symmetries. However, the value for GinUE serves as a good reference mark. We discuss the system size dependence of numerical results in Appendix \ref{app:size_dep}.

\begin{table}[t]
\centering
\renewcommand{\arraystretch}{2}
\begin{tabular}{|>{\centering\arraybackslash}p{2cm}|>{\centering\arraybackslash}p{1.7cm}|>{\centering\arraybackslash}p{1.7cm}|>{\centering\arraybackslash}p{1.7cm}|>{\centering\arraybackslash}p{1.7cm}|>{\centering\arraybackslash}p{1.7cm}|>{\centering\arraybackslash}p{1.7cm}|}
\hline
Value of $q$                    & $e^{i\pi/10.4}$ & $e^{i\pi/8.5}$ & $e^{i\pi/6.7}$ & $e^{i\pi/5.6}$ & $e^{i\pi/4.8}$ & $e^{i\pi/2.7}$ \\ 
\hline
$\mathcal{N}_{\mathbb{C}}$ & 0               & 10             & 335            & 858            & 1464           & 1544           \\ \hline
\end{tabular}
\caption{Maximal number of complex eigenvalues, $\mathcal{N}_{\mathbb{C}}$, that appears for a fixed value of $q$ when $\lambda$ reaches maximally chaotic regime. Total number of eigenvalues in the studied sector is $1688$.}
\label{tbl:max_complex}
\end{table}

To make sure that the chosen sector does not exhibit a unique behaviour, we have also performed analogous computations in the $\{\ell=2,m=2\}$ subsector. The qualitative behaviour of the spectrum in both sectors is similar.

\begin{figure}[t]
    \centering
    \includegraphics[width=1.0\linewidth]{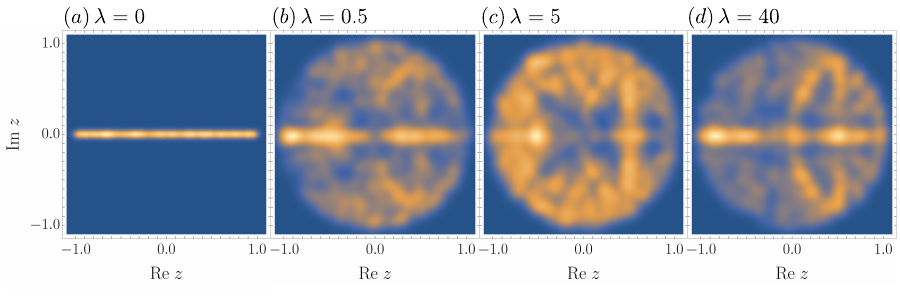}
    \caption{Complex level spacing ratio distributions $P(z)$ for a fixed value of $q=e^{i\pi/4.8}$, fixed length $L=16$, and different values of $\lambda$. Eigenvalues are from $q$-parity even sector with $\{\ell=1,S^z=1\}$. }
    \label{fig:l16q4d8complexRatioDistr}
\end{figure}

Let us also plot the complex level spacing distribution in the complex plane for a fixed value of $q=e^{i\pi/4.8}$. The distributions for different $\lambda$ are presented in the Figure \ref{fig:l16q4d8complexRatioDistr}. As anticipated, for $\lambda=0$ all the eigenvalues are on the real line, as well as the level spacing ratio. For $\lambda>0$ the eigenvalues start to enter the complex plane. For $\lambda=5$ we see that the density $P(z)$ is much smaller around $z=0$, which indicates the level repulsion in the chaotic regime. 

Some of the CLSR plots also exhibit a ``bow and arrow'' feature: an increased density along an ``arc'', a straight vertical line at $\text{Re}\;z = \frac{1}{2}$, and the real line. This feature is explained by PT-symmetry \cite[Appendix D]{Li:2024uzg}, and was also observed in \cite{Akemann:2024unk}. The values of $z_k$ start to form an arc when two eigenvalues coincide and move into a complex plane in the opposite directions, while having the third nearest eigenvalue on the real line. The values of $z_k$ transition from the arc to the vertical line when two complex conjugated eigenvalues start being further away from each other than from the third real one. We make more plots for different values of $q$ to investigate this structure in Appendix \ref{app:CLSR_plots}.

The numerical results for average absolute CLSR, $\braket{r}$, presented in the Figure \ref{fig:average} together with the plot for CLSR in the complex plane given in the Figure \ref{fig:l16q4d8complexRatioDistr} provide a clear evidence that the system \eqref{eq:ham_final} is not integrable. Recall that $\mathcal{H}$ is invariant under $U_q(sl_2)$ for any value of the coupling constant in front of the next-to-nearest interaction term. This provides a strong evidence that $U_q(sl_2)$ global symmetry exists independently of the integrability of the system.

\section{Conclusion}

In this paper we addressed the question whether a quantum group global symmetry can exist independently of integrable structure of the theory. We constructed the spin chain Hamiltonian \eqref{eq:ham_final} with next-to-nearest interaction term invariant under $U_q(sl_2)$ symmetry. We studied the system numerically and found that after projecting out all known global symmetries, including $U_q(sl_2)$, the systems exhibits chaotic behavior . This indicates that there is no other ``hidden'' integrable structure left, and the QG provides finite number of conservation laws, similar to standard Lie groups global symmetries, e.g. $SU(2)$.

The signature of quantum chaotic behavior that we observed is the eigenvalue repulsion within a fully resolved irreducible symmetry sector. The system that we studied is not unitary, instead it possesses PT-symmetry. PT-symmetry constrains the eigenvalues to be either real or to come in complex conjugated pairs. Because of that the analysis of eigenvalue statistics was performed separately for different regimes of parameters of the model. In the case of unbroken PT-symmetry, i.e. when all the eigenvalues are real, the analysis is similar to unitary systems. The level spacing distribution in this case matches with Poisson distribution when the next-to-nearest interaction is turned off and shifts towards Wigner distribution when it is turned on. Distribution shows a good matching with Wigner GOE ensemble for maximally chaotic regime, see Figures \ref{fig:l16q10d4withLocUnfold}, \ref{fig:l16q10d4ratiodistr}. A more generic approach that we use is the statistics of ratios of complex level spacings. It can be applied to the case of so-called broken PT-symmetry. An average absolute value of these distributions serves as a good marker for switching between integrable and chaotic regimes, see Figure \ref{fig:average}. We also show examples of such complex distributions in Figure \ref{fig:l16q4d8complexRatioDistr}.

Even though integrable systems are great models for theoretical framework and non-perturbative calculations, in the real-world physical systems integrability is always violated by some perturbations. From the point of view of lattice systems, which at the moment form the main example of a system with QG symmetry, any external noise would lead to deviations from integrability. The fact that QG can survive, at least under some of the perturbations, strengthens its chances to be a symmetry of some realistic systems. Furthermore, these results open new directions towards the search for QG symmetries in manifestly non-integrable theories. Most likely, such systems will have degeneracies between energies of states created by local and non-local operators, as it happens for non-invertible symmetries \cite{Cordova:2024nux}. Integrability is always broken in interacting relativistic higher-dimensional QFTs due to the Coleman-Mandula theorem which prohibits infinite-dimensional symmetries. The fact that QG is not directly related to integrability, gives a hope that such symmetry can be found in generic systems in higher dimensions. First steps in this direction were made very recently in \cite{Garre-Rubio:2025bdq}.

\section*{Acknowledgements}
We thank Barak Gabai, Igor Klebanov, Andreas Läuchli, Rafael Nepomechie, Bendeguz Offertaler, Jiaxin Qiao, Ruth Shir, Grigory Tarnopolsky, Bernardo Zan for discussions and comments on the draft. AZ would like to thank Elena Petrova for valuable discussions and for suggesting useful references. The work of VG is supported by Simons Foundation grant 994310 (Simons Collaboration on Confinement and QCD Strings).

\appendix

\section{IR phase of the spin chain}
\label{app:IR_effects}
In this appendix we discuss the low energy theory arising in the infinite volume limit of the spin chain \eqref{eq:PS_ham} as well as the effects of adding a next-to-nearest interaction term \eqref{eq:nnInt_pauli}. 

Let us start with a spin chain given by the Hamiltonian \eqref{eq:PS_ham}. As it is common for the antiferromagnetic spin chains of XXZ-type \cite{kadanoff1979correlation,lukyanov1998low,alcaraz1987conformal}, for the choice of the parameter $q$ being on the unit circle \eqref{eq:q_param} the theory flows to a CFT in the IR. For the open spin chain an appropriate field theory description is given in terms of a boundary CFT. The central charge and the spectrum of the theory with Hamiltonian \eqref{eq:PS_ham} were computed in \cite{Pasquier:1989kd} using Bethe ansatz. The effective central charge is given by 
\begin{equation}\label{eq:central_charge}
    c = 1 - \frac{6}{\mu(\mu+1)},
\end{equation}
where $\mu$ is related to $q$ by the formula \eqref{eq:q_param}. One can notice that $c \le 1$ for the allowed values of $\mu \ge 1$. The conformal dimensions of boundary primary operators are given conveniently by dimensions from the Kac table \cite{DiFrancesco:1997nk}, which are labeled by two integers $r,s$:
\begin{equation}\label{Kac_dim}
    h_{r,s} = \frac{[r(\mu+1)-s\mu]^2 - 1}{4\mu(\mu+1)}.
\end{equation}
Let us denote the characters of associated degenerate Virasoro representations as $\chi_{r,s}(\tau)$. Then the partition function of a boundary CFT is given by
\begin{equation}\label{eq:bdry_partfn}
    Z(\tau) = \sum_{j=0}^{\infty} (2j+1) \chi_{1,2j+1}(\tau).
\end{equation}
Here, $(2j+1)$ degeneracy of the spectrum corresponds to the dimensions of irreducible representations of the $U_q(sl_2)$.

\begin{figure}[h]
    \centering
    \includegraphics[width=0.6\linewidth]{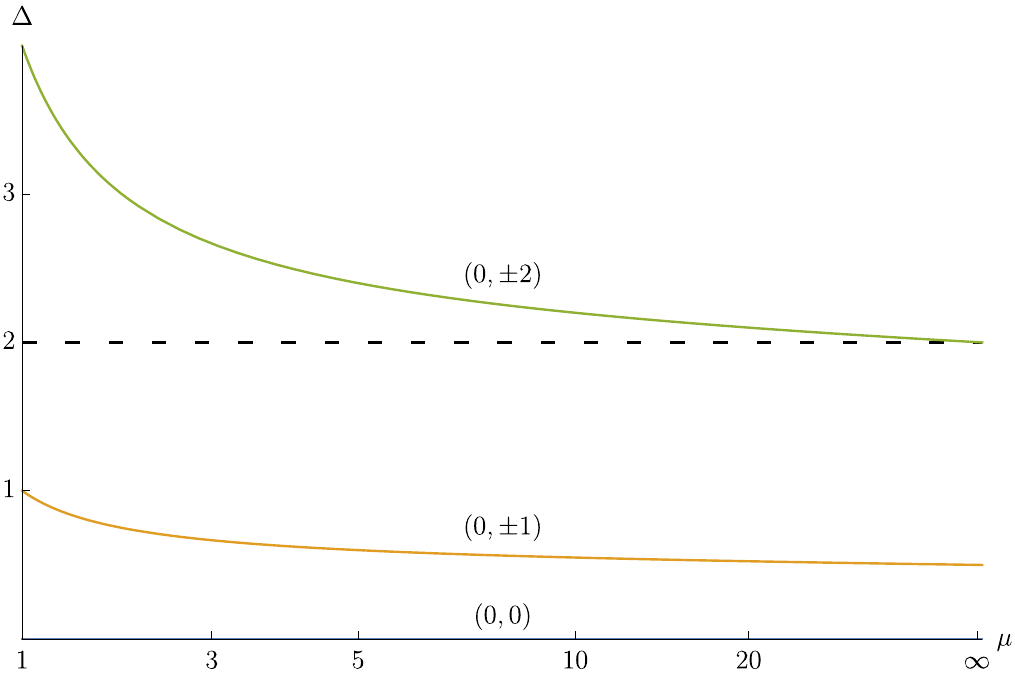}
    \caption{Plot of conformal dimensions $\Delta_{e,m}(\mu)$ for several lightest bulk XXZ operators labeled by quantum numbers $(e,m)$, which are also singlets under the lattice $U(1)_{\text{lat}}$. $\Delta$ is defined as $\Delta = \kappa+\bar{\kappa}$, see \eqref{eq:compBos_dim}. Operators labeled by $(0,\pm 1)$ are relevant for all values of $\mu$, whereas operators $(0,\pm 2)$ are instead irrelevant with approaching marginality at $\mu \to \infty$.
    }
    \label{fig:singlets_spectrum}
\end{figure}

A field theory analysis of the IR phase requires to check that relevant operators are not generated in the RG flow. Bethe ansatz computations of \cite{Pasquier:1989kd} provide a strong evidence that no relevant operators are being generated. Let us comment on why this can happen. The boundary theory \eqref{eq:bdry_partfn} does not have any singlets apart from the identity operator, so one has to consider bulk operators. Naively, the bulk of the Hamiltonian \eqref{eq:PS_ham} is the standard XXZ chain. Local bulk operators in this case are just vertex operators in $c=1$ compact boson theory. The relation between radius $R$ of compactification and $\mu$ is then given by $R = \sqrt{\frac{2(\mu+1)}{\mu}}$, where convention is such that self-dual radius is $R_{\text{s.d.}}=\sqrt{2}$. Vertex operators are labeled by two integers $(e,m)$, quantum numbers under the $U(1)_e \times U(1)_m$ global symmetry. A notation is such that the electric $U(1)_e$ symmetry is the lattice $U(1)_{\text{lat}}$ symmetry. Conformal dimensions of the vertex operators in our conventions are given by
\begin{equation}\label{eq:compBos_dim}
    \kappa_{e,m} = \frac{1}{2} \left( \frac{n}{R} + \frac{mR}{2} \right)^2, \qquad \bar{\kappa}_{e,m} = \frac{1}{2} \left( \frac{n}{R} - \frac{mR}{2} \right)^2
\end{equation}
Singlets under $U(1)_{\text{lat}}$ have quantum numbers $(0,m)$. Let us plot the energy spectrum, $\Delta_{e,m}(\mu)$ (recall that $\Delta=\kappa+\bar{\kappa}$), of several low-lying singlet operators, see Figure \ref{fig:singlets_spectrum}. As one can see from the plot the operators $(0,\pm 1)$ are relevant for all the allowed values of $\mu$. These operators are known not to be generated in the RG flow, see \cite{Cheng:2022sgb} for a recent discussion.\footnote{See also \cite{Affleck:1987ch,tsvelik2007quantum} for a similar discussion in the XXX chain which appears in the $\mu \to \infty$ limit.} The idea is that in the bulk the XXZ Hamiltonian has a lattice translation symmetry $\mathbb{Z}_L$, and in particular, it is invariant also under the lattice translation by one site. However, the one-site translation in the continuum limit becomes an internal symmetry $\mathbb{Z}_2^C \subset U(1)_e \times U(1)_m$. It is given by $e^{i\pi(Q_e + Q_m)}$, where $Q_e$ and $Q_m$ are correspondingly electric, $U(1)_e$, and magnetic, $U(1)_m$, charges. It is clear then that the operators $(0,\pm 1)$ are odd under this $\mathbb{Z}_2^C$ symmetry and it protects them from being generated in the RG flow.\footnote{On the one hand, one cannot call $\mathbb{Z}_2^C$ symmetry an internal symmetry on the lattice because it does not preserve local operators. On the other hand, it is also not an emergent symmetry of the IR phase, because it is not violated by any irrelevant operator generated by the Hamiltonian. In \cite{Cheng:2022sgb} it is suggested to call such a symmetry as emanant.} The next after the lightest are the vertex operators $(0,\pm2)$, which are irrelevant for all $\mu \ge 1$. In the limit $\mu \to \infty$ their dimension $\Delta \to 2$, which makes it possible to construct a well known marginally irrelevant operator \cite{Affleck:1988px,affleck1998exact,Lukyanov:2002fg} leading to logarithmic corrections.

In the discussion of the bulk RG flow we did not use QG symmetry of the system, and indeed the spectrum of local bulk operators displayed in the Fig. \ref{fig:singlets_spectrum} is not organized in the multiplets of $U_q(sl_2)$. The reason for this is that this spectrum corresponds to the IR limit of the standard XXZ spin chain with closed boundary conditions that break QG symmetry. It is natural to expect that our open spin chain also contains defect-ending operators in the bulk that transform in representations of $U_q(sl_2)$. Likely, these operators have dimensions similar to those of the XXZ$_q$ CFT studied in \cite{paper1}. It would be nice to understand which constraints QG imposes on the bulk RG flow of our system. In any case, the spectrum of XXZ$_q$ CFT contains a relevant $U_q(sl_2)$ singlet, and hence the $\mathbb{Z}_2^C$ symmetry discussion presented above is in any case necessary to explain the stability of the IR phase.

We also perform numerical computations that support the appearance of a CFT in the IR. If the spin chain flows to a CFT, for the low-energy states of open chains the scaling behavior is known to be \cite{Alcaraz:1987zr}
\begin{equation}\label{eq:excited_states}
    E_i^\mu = E_0^\mu(L) + \frac{\pi\zeta^\mu}{L} (\Delta_i^\mu + \mathbb{N}) + O\left(\frac{1}{L^2}\right),
\end{equation}
where $E_0^\mu(L)$ is the ground state energy\footnote{For the open spin chains the scaling behavior of the ground state energy is given by
\begin{equation}
    E_{0}^{\mu}(L) = L e_{\infty}^{\mu} + f_{\infty}^{\mu} - \frac{\pi \zeta^{\mu} c}{24L} + O \left(\frac{1}{L^{2}} \right),
\end{equation}
where the $O(L)$ and $O(1/L)$ terms are similar to those in the closed spin chains: $e_\infty^\mu$ is the vacuum energy density; $c$ is the central charge \eqref{eq:central_charge}; $\zeta^\mu$ is the speed of sound. The $O(1)$ term, $f_\infty^\mu$, is the new one and encodes the surface density. For the spin chain \eqref{eq:PS_ham} the values of $f_\infty^\mu$ were found in \cite{Alcaraz:1987uk} by solving the Bethe ansatz equations numerically. For our analysis of excited states \eqref{eq:excited_states} we subtract the ground state energy $E_0^\mu(L)$, thus we do not need to separately find $f_\infty^\mu$.}; $\zeta^\mu$ is the so called speed of sound in the model; $\Delta_i$ is the conformal dimension of primary operator and $\mathbb{N}$ denotes the Virasoro descendants which come at integer level spacing in such a normalization. 

\begin{figure}[t]
    \subfloat{{\includegraphics[width=5cm,valign=c]{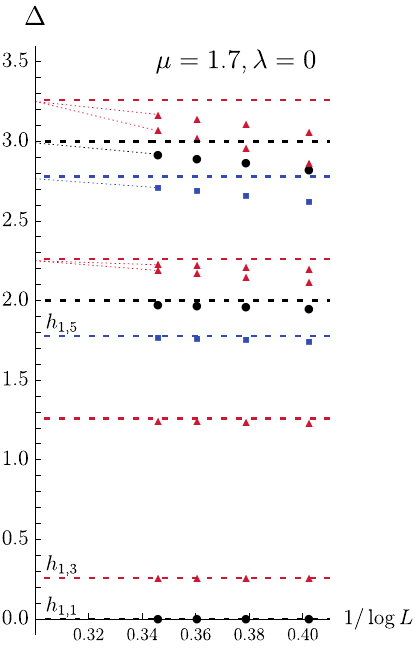} }}
    \quad
    \subfloat{{\includegraphics[width=5cm,valign=c]{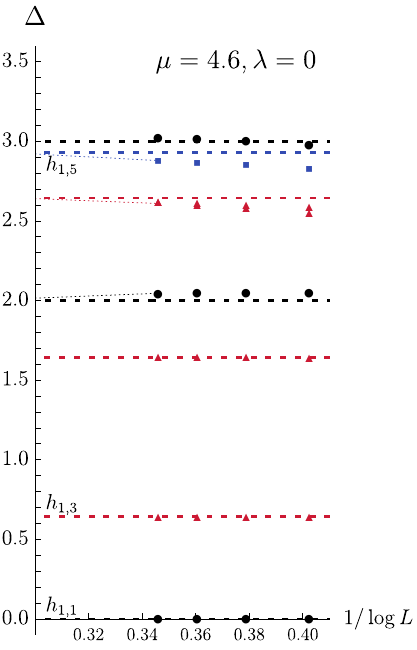} }}
    \quad
    \subfloat{{\includegraphics[width=5cm,valign=c]{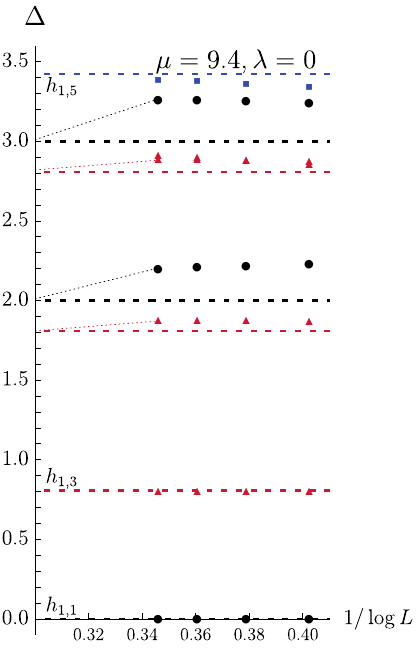} }}
    \caption{Each figure shows the first few scaling dimensions $\Delta_i$ of the unperturbed Hamiltonian \eqref{eq:PS_ham} in the sector $S^z=0$ for spin chain length $L = \{12,14,16,18\}$. The parameter $\mu$, \eqref{eq:q_param}, changes between the figures. The dashed lines represent the exact conformal dimensions from the spectrum \eqref{eq:bdry_partfn}. Color coding is such that all the states in the Virasoro multiplet of $\mathbb{1}$ are black, in the multiplet of $h_{1,3}$ are red, and in the one of $h_{1,5}$ are blue. The ground state and the first excited state always have assigned values $0$ and $h_{1,3}$ respectively. The other eigenvalues form Virasoro towers of states with the increase of $L$. The dotted lines represent the expected extrapolation of eigenvalues for the infinite length chain.}
    \label{fig:IntSpacing_0}
\end{figure}
We use exact diagonalization to study numerically the several lowest eigenvalues of the Hamiltonian \eqref{eq:PS_ham} for different length of the spin chain $L=\{12, 14, 16, 18\}$, as well as for different values of the parameter $\mu$, \eqref{eq:q_param}. Figure \ref{fig:IntSpacing_0} represents the results of numerical diagonalization of the Hamiltonian in a given sector with $S^z=0$. This sector contains the ground state, all the primary states with degeneracy 1, as well as all their Virasoro descendants. The quantum group degeneracies appear when considering sectors with different quantum number $S^z$. Our aim is to measure the conformal dimensions, $\Delta_i$-s, and observe an appearance of the Virasoro tower of states, so that for our purposes it is enough to consider only the sector with $S^z=0$. We also perform the computations for several values of $\mu$. Scaling dimensions in the Figure \ref{fig:IntSpacing_0} are obtained using the formula \eqref{eq:excited_states} in the following way. The ground state energy is always subtracted to be 0. To get the excited states one then has to normalize the eigenvalues with a coefficient $\frac{L}{\pi \zeta^\mu}$, which contains the speed of sound. For an infinite chain $\zeta^\mu$ can be computed using Bethe ansatz, however, for a finite chain one can normalize the first eigenvalue to be the first excited state in the spectrum \eqref{eq:bdry_partfn}, that is $h_{1,3}$. See for example \cite{Lauchli:2013jga,Huang:2023hqx} for a similar approach. After such a normalization one can see that as $L$ increases the eigenvalues organise themselves into several Virasoro towers with primary operators given by $h_{1,2j+1}$, according to the spectrum \eqref{eq:bdry_partfn}. In particular, one can notice that the number of descendants is correct for each degenerate Virasoro multiplet. For the multiplet of $\mathbb{1}$ there is no descendant on level 1, and a single descendant on levels 2 and 3. Multiplet of $h_{1,3}$ has first null state on level 3, so that there are only 2 descendants on this level. The convergence is better for smaller values of $\mu$, which is consistent with irrelevant operators $(0,\pm2)$ moving away from $\Delta=2$ with decrease of $\mu$, see Figure \ref{fig:singlets_spectrum}.

\begin{figure}[t]
    \subfloat{{\includegraphics[width=5cm,valign=c]{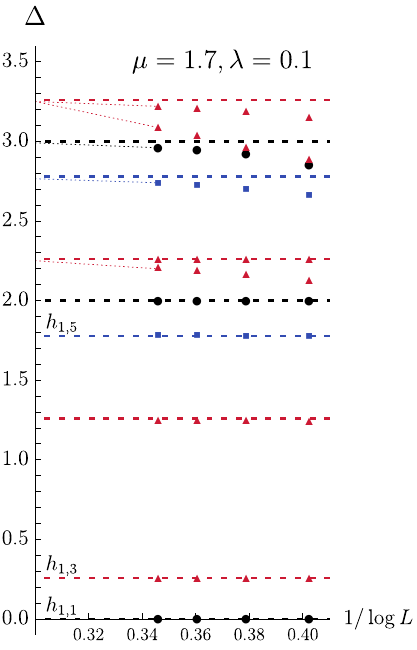} }}
    \quad
    \subfloat{{\includegraphics[width=5cm,valign=c]{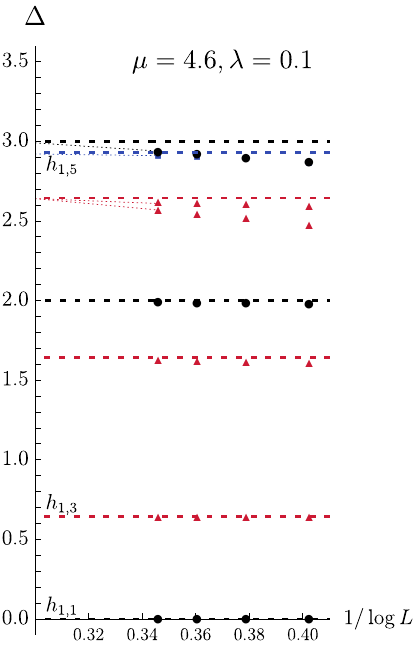} }}
    \quad
    \subfloat{{\includegraphics[width=5cm,valign=c]{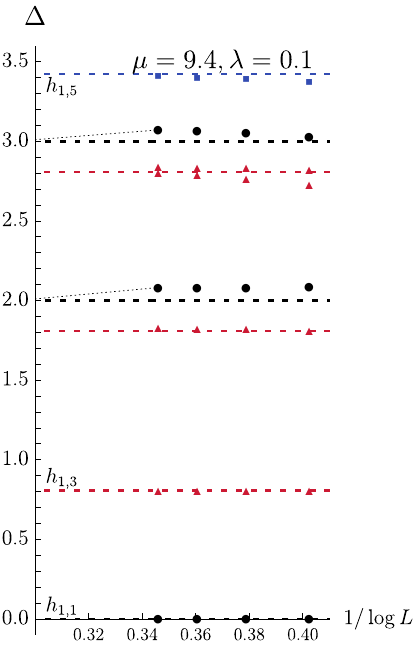} }}
    \caption{Each figure shows the first few scaling dimensions $\Delta_i$ of the Hamiltonian \eqref{eq:ham_final}, perturbed by a next-to-nearest interaction term with a small coupling constant $\lambda=0.1$. Eigenvalues are computed in the sector $S^z=0$ for spin chain of length $L = \{12,14,16,18\}$. The construction of the plots is the same as in the Figure \ref{fig:IntSpacing_0}.}
    \label{fig:IntSpacing_pert}
\end{figure}
Let us now return to the Hamiltonian \eqref{eq:ham_final}, perturbed by a next-to-nearest interaction term. When the coupling $\lambda$ is small, we can study it as a perturbation of the IR CFT. Interaction term \eqref{eq:nnInt_pauli} preserves all the symmetries of the original Hamiltonian. Given the discussion above for the unperturbed case, it suggests that the perturbation term is irrelevant in the RG sense. So that in the IR we expect the deformed theory flow to the same CFT \eqref{eq:bdry_partfn}. The exact diagonalization of the perturbed Hamiltonian supports this statement, see Figure \ref{fig:IntSpacing_pert}. Similarly to the previous figure, we plot the first few eigenvalues of the Hamiltonian \eqref{eq:ham_final} with a small value of the coupling constant, $\lambda=0.1$.  We observe that with the increase of $L$ the states form the same Virasoro towers of states.

\section{Complex level spacing ratio distributions}
\label{app:CLSR_plots}
\begin{figure}[t]
    \centering
    \begin{subfigure}{\textwidth}
        \centering
        \includegraphics[width=1.0\linewidth]{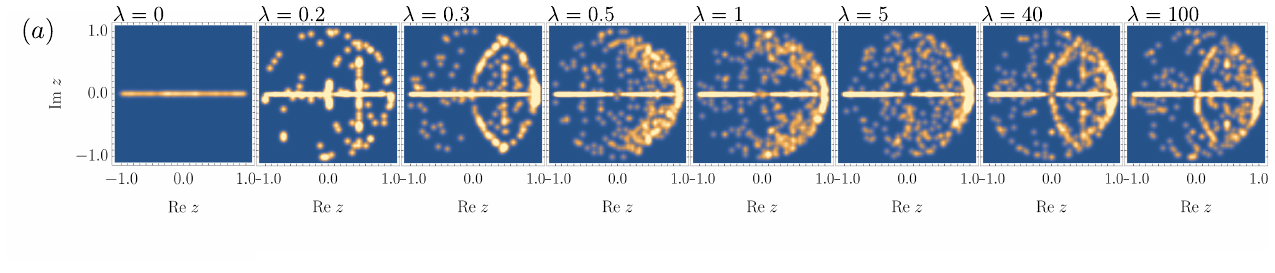}
    \end{subfigure}

    \begin{subfigure}{\textwidth}
        \centering
        \includegraphics[width=1.0\linewidth]{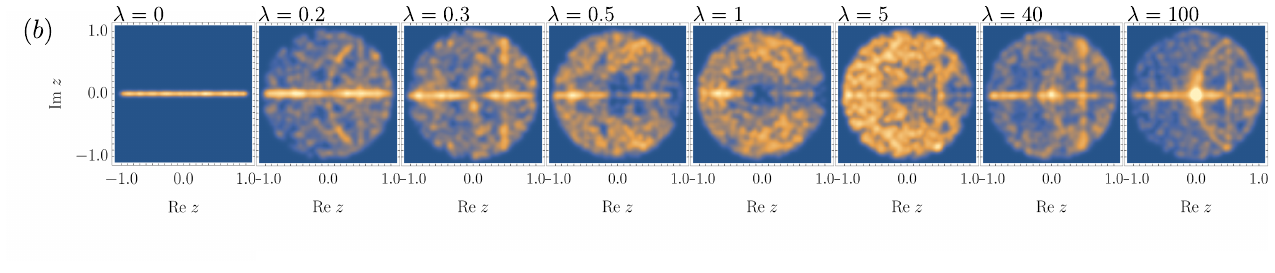}
    \end{subfigure}

    \caption{Complex level spacing ratio distributions $P(z)$ for different values of $\lambda$. Parameter $q$ is: (a) $q=e^{i\pi/6.7}$; (b) $q=e^{i\pi/2.7}$. Spin chain length is $L=16$. Eigenvalues are from $q$-parity even sector with $\{\ell=1,S^z=1\}$. }
    \label{fig:clsr_app}
\end{figure}

In this appendix we present more numerical results of complex level spacing ratio (CLSR) distributions $P(z)$ for different values of parameter $q$ and for different values of coupling $\lambda$, see Figure \ref{fig:clsr_app}. Part (a) of the figure shows CLSR for $q=e^{i\pi/6.7}$; part (b) shows CLSR for $q=e^{i\pi/2.7}$, this is in addition to $q=e^{i\pi/4.8}$ shown in the main text. The difference between these values of the parameter $q$ is the number of eigenvalues that become complex with the increase of $\lambda$. In (a) the maximal number of complex eigenvalues reaches 335 out of 1688, that is, the majority of eigenvalues stays real for any value of $\lambda$, see Table \ref{tbl:max_complex}. In (b) this number is 1544 out of 1688, which makes almost all eigenvalues to be complex in the most chaotic regime. Moreover, in (b) the PT-symmetry breaking happens for very small values of $\lambda$, which means that for all plots presented in (b) the majority of eigenvalues is complex.

Despite having different values of $q$ and therefore different number of complex eigenvalues, both (a) and (b) exhibit the ``bow and arrow'' structure, an artifact of PT symmetry (see discussion around Figure \ref{fig:l16q4d8complexRatioDistr}). Such a structure is best visible both for small and very large values of the deformation parameter $\lambda$. This structure emerges when there are enough triples of eigenvalues close to each other, one real and two complex conjugated ones. Interestingly, this structure disappears or becomes indistinguishable from the rest of the distribution for the most chaotic regime, where the values of $\lambda$ are between $0.5$ and $5$. It would be interesting to investigate further any correlation between ``bow and arrow'' structure and integrability-chaos transition.

\section{System size dependence}
\label{app:size_dep}
\begin{figure}[h]
    \centering
    \includegraphics[width=0.7\linewidth]{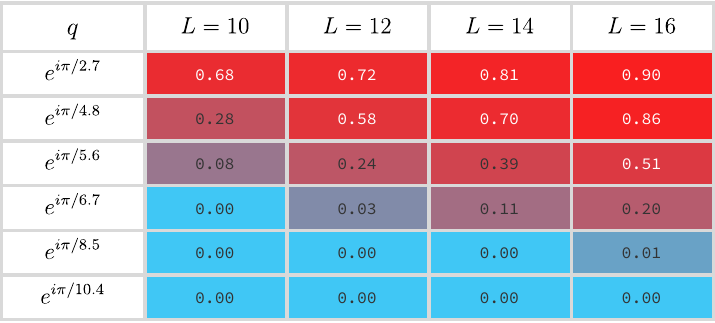}
    \caption{The table represents which fraction of eigenvalues is complex for different values of $q$ and $L$. Eigenvalues are taken from the $q$-parity even sector with $\{\ell=1,m=1\}$. Table cells are color-coded on a blue-to-red gradient, where blue represents fraction$=$0 and red represents 1.}
    \label{fig:complex_fraction}
\end{figure}

In this appendix we would like to address the dependence of numerical results on the system size $L$. First, we would like to discuss the appearance of complex eigenvalues (PT-symmetry breaking). Let us choose a fixed value of $\lambda$ to be in a chaotic regime, for example $\lambda=5$. Now, for all studied values of $q$, and for $L=10,12,14,16$
we calculate  which fraction of eigenvalues is complex. We remind the reader that we work with the $q$-parity even sector with quantum numbers $\{\ell=1,m=1\}$, the total number of eigenvalues in this sector for mentioned values of $L$ is 50, 141, 518, 1688 respectively. We calculate the number of complex eigenvalues for each value of the parameters and divide respectively by the total number of eigenvalues. The results are presented in Figure \ref{fig:complex_fraction}. As one can see, with the increase of the system size complex eigenvalues appear for values of $q$ closer and closer to 1. Extrapolating the given results, the observed trend suggests that as $L \to \infty$ complex eigenvalues would appear for any $q \ne 1$. We also do not see any indication of existence of some critical value $q^*$ other than $q=1$ for which the PT-symmetry breaking phenomenon can happen.

\begin{figure}[h]
    \centering
    \includegraphics[width=0.6\linewidth]{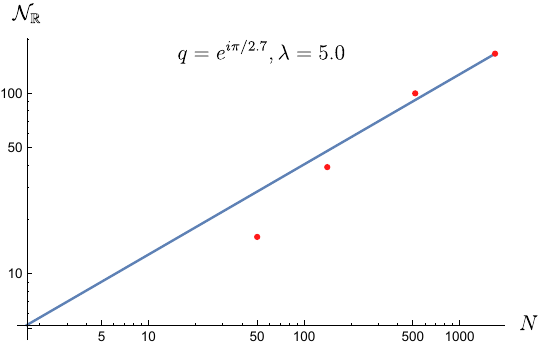}
    \caption{The figure shows the log-log plot of a number of real eigenvalues in the spectrum as a function of the matrix size $N$. Each red dot represent the data for different system sizes $L=10,12,14,16$. Matrices are studied for parameters $q=e^{i\pi/2.7}, \lambda=5.0$ in the sector with quantum numbers $\{\ell=1,m=1\}$. The fitted value $\epsilon \simeq 0.25$.}
    \label{fig:sqrt_fit}
\end{figure}

Let us now study how well the ensemble converged to its asymptotic prediction for a given system size. To do it we study the number of real eigenvalues in the spectrum and compare it with asymptotic predictions. We expect the number of real eigenvalues to behave as $\mathcal{N}_{\mathbb{R}} \sim \sqrt{N}/\epsilon$ in the asymptotic regime, where $N$ is the size of the matrix, and $\epsilon$ is a parameter controlling the Hermiticity breaking which is small for $q\approx 1$. For the most complex studied value of $q$, $q=e^{i\pi/2.7}$, Figure \ref{fig:sqrt_fit} shows the plot of number of real eigenvalues versus the size of the matrix $N$ for different values of $L=10,12,14,16$. The fitted value of $\epsilon \simeq 0.25$. We see that for system sizes $L=14,16$ the number of real eigenvalues has converged to the square root formula. For the other studied values of $q$ closer to 1, the expectation is that more complex eigenvalues will appear with the increase of the system size, until it reaches values close to $N$ with corrections of order $N^{-1/2}$. 

We thus conclude that while behavior of Eigenvalues is already chaotic for all $q$'s we studied for $\lambda$'s in the appropriate range, an ensemble which governs their statistics is already GinUE for most complex $q$, still GOE for $q$'s closest to 1, and is  transitioning between the two regimes for intermediate $q$'s for the largest values of $N$ we could achieve.

\bibliography{biblio_eigenv}
\bibliographystyle{utphys}
\end{document}